\documentclass[10pt]{article}
\usepackage{times}

\def\fd#1#2{\frac{d#1}{d#2}}
\def\pd#1#2{\frac{\partial#1}{\partial#2}}
\def\pdd#1#2#3{\ifx#2#3\frac{\partial^2#1}{\partial#2^2}\else\frac{\partial^2#1}{\partial#2\mkern1mu\partial#3}\fi}

\let\tint=\int
\def\eqnitem#1. {\par\medbreak\leavevmode#1$^\circ$.\enspace}
\def\Example#1.{{\sl Example\hskip.5em\relax#1\unskip.}}
\def\Remark#1.{{\sl Remark\hskip.5em\relax#1\unskip.}}
\begin{document}

\title{The von Mises transformation: order reduction and construction of
B\"acklund transformations and new integrable equations}

\author{Andrei D.~Polyanin\footnote{Institute for Problems in Mechanics, Russian Academy of Sciences,
  101 Vernadsky Avenue, bldg 1, 119526 Moscow, Russia. E-mail: polyanin@ipmnet.ru} \and
Alexei I.~Zhurov\footnote{Institute for Problems in Mechanics, Russian Academy
of Sciences, 101 Vernadsky Avenue, bldg 1, 119526 Moscow, Russia; School of
Dentistry, Cardiff University, Heath Park, Cardiff CF14 4XY, UK. \hbox{E-mail}:
zhurovai@cardiff.ac.uk}}
\date{Original version 3 July 2009, revised 17 July 2009}
\maketitle

\begin{abstract}
Wide classes of nonlinear mathematical physics equations are described that
admit order reduction through the use of the von Mises transformation, with
the unknown function taken as the new independent variable and an
appropriate partial derivative taken as the new dependent variable.
RF-pairs and associated B\"acklund transformations are constructed for
evolution equations of general form (special cases of which are
Burgers, Korteweg--de Vries, and Harry Dym type equations as well as many
other nonlinear equations of mathematical physics). The results obtained
are used for order reduction and constructing exact solutions of
hydrodynamics equations. A generalized Calogero equation and a number of
other new integrable nonlinear equations are considered.

\end{abstract}

\section{Preliminary remarks}

A simple analogue of the von Mises transformation is widely used in
the theory of ordinary differential equations. Specifically,
an $n$th-order autonomous ordinary differential equation of general form,
\begin{equation}
F(u,u'_x,u''_{xx},\dots,u^{(n)}_x)=0,
\label{(1.1)}
\end{equation}
which does not depend explicitly on $x$, can be reduced
with the change of variable
\begin{equation}
\eta(u)=u'_x\qquad
(\mbox{\it analogue of the von Mises transformation})
\label{(1.2)}
\end{equation}
to an $(n-1)$st-order equation (e.g., see [1, 2]).

For partial differential equations, the von Mises transformation (with the
unknown function and its appropriate partial derivative taken as the new
independent and dependent variables, respectively) is applied for order
reduction of steady hydrodynamic boundary layer equations~[3--5].

An RF-pair is an operation of consecutive raising and lowering of the order
of an equation (RF stands for ``Rise'' and ``Fall''). To illustrate this concept,
let us solve the $n$th-order ordinary differential equation of general form
\begin{equation}
F(x,u,u'_x,u''_{xx},\dots,u^{(n)}_x)=0
\label{(1.3)}
\end{equation}
for the independent variable (which is assumed to be feasible) to obtain
\begin{equation}
G(u,u'_x,u''_{xx},\dots,u^{(n)}_x)=x.
\label{(1.4)}
\end{equation}
Differentiating (\ref{(1.4)}) with respect to~$x$ yields
an $(n+1)$st-order autonomous equation, whose order is further
reduced using substitution~(\ref{(1.2)}). As a result, we arrive at
the $n$th-order equation
\begin{equation}
\eta\fd{}u G(u,\eta,\eta\eta_u',\dots,\eta^{(n-1)}_x)=1.
\label{(1.5)}
\end{equation}
The derivatives with respect to $x$ are found consecutively
using $\eta^{(k)}_x=\eta(\eta^{(k-1)}_x)_u$. The derivative with respect to~$u$
on the left-hand side of~(\ref{(1.5)}) is calculated by the chain rule taking
into account that $F$ depends on $u$,~$\eta$, $\eta_u$,~\dots

Equations (\ref{(1.4)}) and (\ref{(1.5)}) are related by
the B\"acklund transformation
\begin{equation}
\begin{array}{rcl}
G(u,\eta,\eta'_{x},\dots,\eta^{(n-1)}_x)&=&x,\\
u'_x&=&\eta.
\end{array}
\label{(1.6)}
\end{equation}

Many new integrable ordinary differential equations have been found in
a systematic manner using RF-pair operations and associated B\"acklund
transformations~[2,~6].

The above ideas and results for ordinary differential equations turn out to be
transferrable, when appropriately modified, to nonlinear partial differential
equations. The present paper will describe wide classes of nonlinear
mathematical physics equations that admit order reduction with the Mises
transformation. This transformation will then be used to construct RF-pairs and
B\"acklund transformations for nonlinear evolution equations of general form.
This will result in finding new integrable nonlinear equations.

Various B\"acklund transformations and their applications to specific
mathematical physics equations can be found, for example, in [5, 7--17].

In the present paper, the term {\it integrable
equation\/} applies to nonlinear partial differential equations that admit
solution in terms of quadratures or can be linearized (i.e., their solutions
can be expressed in terms of solutions to linear differential equations or
linear integral equations).

\section{Using the von Mises transformation for order\\ reduction of partial differential equations}

Consider the nonlinear $n$th-order equation [5, p.~678]
\begin{equation}\label{(2.1)}
\pd ux\pdd utx-\pd ut\pdd uxx=F\left(t,u,\pd ux,\pdd uxx,\ldots,\pd{^nu}{x^n}\right).
\end{equation}
In the special case $F=\nu u_{xxx}$, this is a nonlinear third-order equation
that describes a laminar boundary layer of a non-Newtonian fluid on a flat
plate, with $u$ being the stream function, $t$~and~$x$ spatial coordinates
measured along and across the plate, and $\nu$ the
kinematic fluid viscosity [3--5].

\eqnitem 1. Characteristic property of Eq.~(\ref{(2.1)}). Let $\widetilde
u(t,x)$ be a solution to the above equation. Then the function
$$
u(t,x)=\widetilde u\left(t,\,x+\varphi(t)\right),
$$
with an arbitrary function $\varphi(t)$, is also a solution to the equation.

\eqnitem 2. In Eq.~(\ref{(2.1)}), let us pass from the old variables to
the Mises variables:
\begin{equation}\label{(2.2)}
t, \ x, \  u=u(t,x)\quad \Longrightarrow\quad
t, \ u, \ \eta=\eta(t,u),\ \ \hbox{where}\ \ \eta=\pd ux.
\end{equation}
The derivatives are transformed as follows:
\begin{equation}\label{(2.3a)}
\pd{}x=\pd u x\pd{} u=\eta\pd{} u,\quad
\pd{}t=\pd{}t+\pd u t\pd{} u,
\end{equation}
\begin{equation}\label{(2.3b)}
\pdd uxx=\eta\pd \eta u, \quad
\pd{^3u}{x^3}=\eta\pd{}u\left(\eta\pd \eta u\right), \quad \dots;\quad
\pdd utx=\pd{\eta}t+\pd u t\pd{\eta} u.
\end{equation}
As a result, the von Mises transformation (\ref{(2.2)}) reduces the $n$th-order
equation (\ref{(2.1)}) to the $(n-1)$st-order equation
\begin{equation}\label{(2.4)}
\eta\pd \eta t=F\left(t,u,\eta,\eta\pd \eta u,\ldots,\pd{^{n-1}\eta}{x^{n-1}}\right),
\end{equation}
where the higher derivatives are calculated as
$$
\pd{^{k} u}{x^{k}}=\pd{^{k-1}\eta}{x^{k-1}}=\eta\pd{}u\pd{^{k-2}\eta}{x^{k-2}},\quad
\pd{}x=\eta\pd{}u,\quad k=3,\dots,n.
$$

Short notation for partial derivatives (e.g., $u_x$, $u_{xx}$, $u_{xt}$, etc.)\
will sometimes be used in what follows.

\medbreak
\Example 1. Equation (\ref{(2.1)}) with
$F=f(t,u)u_{xx}+g(t,u)u_x^2+h(t,u)u_x=f(t,u)\eta\eta_u+g(t,u)\eta^2+h(t,u)\eta$
is reduced to an equation (\ref{(2.4)}) of a special form, which is further
reduced, after canceling by~$\eta$, to a linear first-order equation:
$$
\eta_t-f(t,u)\eta_u=g(t,u)\eta+h(t,u).
$$

\Example 2. If $F=f(t,u)u_{xxx}/u_x=f(t,u)(\eta\eta_u)_u$, Eq.~(\ref{(2.4)}) is
reduced with the change of variable $Z=\eta^2$ to a linear second-order
equation: $Z_t=f(t,u)Z_{uu}$.

\medbreak Given below are two possible generalizations of Eq.~(\ref{(2.1)}).

\eqnitem 1. Equation (\ref{(2.1)}) can be generalized by including $I_x$,
\dots, $I_x^{(m)}$, where $I=u_xu_{tx}-u_tu_{xx}$, into~$F$ as additional
arguments.

\eqnitem 2. The von Mises transformation (\ref{(2.2)}) reduces the order of the
equation
\begin{equation}\label{(2.5)}
u_xG_t-u_tG_x=F(t,u,u_x,\ldots,u_x^{(n)}),\quad G=G(t,u,u_x,\ldots,u_x^{(m)}),
\end{equation}
and brings it to the form (\ref{(2.4)}), where $\eta_t$ must be replaced
with~$G_t$. If $G=u_x$, Eq.~(\ref{(2.5)}) coincides with Eq.~(\ref{(2.1)}).

\medbreak
\Example 3. The nonlinear third-order equation
\begin{equation}
u_xu_{txx}-u_tu_{xxx}=f(t,u)u_x
\label{(2.5a)}
\end{equation}
is a special case of Eq.~(\ref{(2.5)}) with $G=u_{xx}$ and $F=f(t,u)u_x$.
The von Mises transformation (\ref{(2.2)}) brings it to the second-order equation
$(\eta\eta_u)_t=f(t,u)$, whose general solution is given by
$$
\eta^2=2\tint^t_{t_0}\!\!\tint^u_{u_0}f(\tau,\xi)\,d\xi\,d\tau+\varphi(t)+\psi(u),
$$
where $\varphi(t)$ and $\psi(u)$ arbitrary functions and $t_0$ and $u_0$ are
arbitrary constants.

\Remark. Other partial differential equations that admit order reduction will
be considered below in Section~\ref{sec:6}.

\section{Using the von Mises transformation for constructing RF-pairs and B\"acklund transformations}

Consider the $n$th-order evolution equation of general form
\begin{equation}\label{(3.1)}
u_t=s(t)xu_x+F(t,u,u_x,u_{xx},\dots,u^{(n)}_x).
\end{equation}
The Burgers, Korteweg--de Vries, and Harry Dym equations as well as many other
nonlinear mathematical physics equations [5, 7--9] are special cases of
Eq.~(\ref{(3.1)}).

\eqnitem 1. Characteristic property of Eq.~(\ref{(3.1)}). Let $\widetilde u(t,x)$ be
a solution to the above equation. Then the function
$$
u=\widetilde u(t,\,x+\psi(t)),\quad \psi(t)=C\exp\left(-\tint s(t)\,dt\right),
$$
where $C$ is an arbitrary constant, is also a solution to this equation.

\eqnitem 2.
Suitable RF-pair operations are performed using the formula
\begin{equation}\label{(3.2)}
\pd{}x\left(\frac{u_t}{u_x}\right)\equiv\frac{u_xu_{tx}-u_tu_{xx}}{u_x^2}=\frac{\eta_t}{\eta}.
\end{equation}
It is obtained by differentiating the ratio of the derivatives $u_t/u_x$ with respect to~$x$
followed by passing from $t$,~$x$,~$u$ to the von Mises variables (\ref{(2.2)}).

Let us divide both sides of Eq.~(\ref{(3.1)}) by $u_x$ and differentiate with
respect to~$x$. The left-hand side of the resulting $(n+1)$st-order equation
coincides with~(\ref{(3.2)}). By passing to the von Mises variables
(\ref{(2.2)}) and rearranging, we arrive at the $n$th-order equation
\begin{equation}\label{(3.3)}
\eta_t=s(t)\eta+\eta^2\pd{}u\left(\frac F\eta\right), \quad
F=F(t,u,\eta,\eta\eta_u,\dots,\eta^{(n-1)}_x).
\end{equation}
The partial derivatives with respect to~$x$ are calculated successively
using $\eta^{(k)}_x=\eta(\eta^{(k-1)}_x)_u$. The derivative with respect to~$u$
on the right-hand side of Eq.~(\ref{(3.3)}) is obtained by the chain rule
taking into account that $F$ depends on $u$,~$\eta$, $\eta_u$,~\dots

\eqnitem 3.
Equations (\ref{(3.1)}) and (\ref{(3.3)}) are related by the B\"acklund transformation
\begin{equation}
\begin{array}{rcl}
u_t&=&s(t)x\eta+F(t,u,\eta,\eta_x,\dots,\eta^{(n-1)}_x),\\
u_x&=&\eta.
\end{array}
\label{(3.4)}
\end{equation}

Any solution $u=g(t,x)$ to Eq.~(\ref{(3.1)}) determines a solution to
Eq.~(\ref{(3.3)}), which can be represented in parametric form as
$u=g(t,x)$, $\eta=g_x(t,x)$.

Suppose $\eta=\eta(t,u)$ is a solution to Eq.~(\ref{(3.3)}). Then it follows
from the second equation in (\ref{(3.4)}) that
\begin{equation}
\int\frac{du}{\eta(t,u)}=x+\varphi(t),
\label{(3.5)}
\end{equation}
where $\varphi(t)$ is an arbitrary function. Further, on solving
Eq.~(\ref{(3.5)}) for $u=u(t,x)$, one should substitute the resulting
expression into the original equation (\ref{(3.1)}) and determine~$\varphi(t)$.

\Remark.
It is sometimes convenient to rewrite Eq.~(\ref{(3.3)}) in the form
$$
\zeta_t=-s(t)\zeta-\pd{}u(\zeta F), \quad
\zeta=\frac 1\eta.
$$

\section{Using the von Mises transformation for constructing new integrable nonlinear equations}

Below are examples of using the B\"acklund transformation (\ref{(3.4)}) to
construct new integrable nonlinear equations of mathematical physics.

\medbreak
\Example 1. Consider the generalized Burgers equation
\begin{equation}
u_t+f(u)u_x=au_{xx}.
\label{(4.1)}
\end{equation}
The B\"acklund transformation (\ref{(3.4)}) with
$F=a\eta_x-f(u)\eta=a\eta\eta_{u}-f(u)\eta$ and $s(t)\equiv 0$
brings Eq.~(\ref{(4.1)}) to the equation $\eta_t=a\eta^2\eta_{uu}-f'_u(u)\eta^2$,
which is further reduced with the substitution
$\eta=1/\zeta$ to the nonlinear heat equation with a source
\begin{equation}
\zeta_t=a(\zeta^{-2}\zeta_u)_u+f'_u(u).
\label{(4.2)}
\end{equation}
By setting $f(u)=bu$ in (\ref{(4.1)}), we arrive at an integrable
equation (\ref{(4.2)}) (reducible to the classical Burgers equation (\ref{(4.1)})),
which was derived in~[18] (see also [5,~19]) using
a different method.

\medskip
\Example 2.
Consider the nonlinear second-order equation
\begin{equation}
u_t=u_{xx}+f(u)u_x^2.
\label{(4.3)}
\end{equation}
It is reducible, with the change of variable [5, p.~87]
$$
w=\int F(u)\,du,\quad \ F(u)=\exp\left[\tint f(u)\,du\right],
$$
to the linear heat equation $w_t=w_{xx}$.

The B\"acklund transformation (\ref{(3.4)}) with
$F=\eta_x+f(u)\eta^2=\eta\eta_{u}+f(u)\eta^2$ and $s(t)\equiv 0$ reduces Eq.~(\ref{(4.3)})
to the new integrable nonlinear equation
$$
\eta_t=\eta^2\eta_{uu}+\eta^2[f(u)\eta]_u.
$$

\medskip
\Example 3.
The linear third-order equation
$$
u_t=au_{xxx}+bu_{xx}+sxu_x
$$
with arbitrary functional coefficients $a=a(t)$, $b=b(t)$, and $s=s(t)$
can be reduced, using the B\"acklund transformation (\ref{(3.4)}) with
$F=a\eta_{xx}+b\eta_{x}= a\eta(\eta\eta_u)_u+b\eta\eta_u$ followed
by substituting $\eta=1/\zeta$, to a new integrable equation
\begin{equation}
\zeta_t+s\zeta=a(\zeta^{-3}\zeta_u)_{uu}+b(\zeta^{-2}\zeta_u)_u.
\label{(4.4)}
\end{equation}
In the special cases $a=s=0$ and $b=s=0$, we have
simpler nonlinear equations, which were studied in~[18] and [5, p.~26, 534].

\medskip
\Example 4.
Consider the nonlinear third-order equation
\begin{equation}
u_t+f(u)u_x=au_{xxx}.
\label{(4.5)}
\end{equation}
If $f(u)=\hbox{const}$, it is linear. For $f(u)=bu$ and
$f(u)=bu^2$, it becomes the Korteweg--de Vries and modified Korteweg--de Vries
equation [5,~7], respectively.
The B\"acklund transformation (\ref{(3.4)}) with
$F=a\eta_{xx}-f(u)\eta=a\eta(\eta\eta_u)_u-f(u)\eta$ and
$s(t)\equiv 0$ brings Eq.~(\ref{(4.5)}) to the equation
$\eta_t=a\eta^2(\eta\eta_u)_{uu}-f'_u(u)\eta^2$, which is further reduced with
the change of variable $\eta=1/\zeta$ to the form
\begin{equation}
\zeta_t=a(\zeta^{-3}\zeta_u)_{uu}+f'_u(u).
\label{(4.6)}
\end{equation}
The special case of this equation with $f(u)\equiv 0$ was obtained
in [5, p.~534] by a different method.

Also, setting $f(u)=bu$ and $f(u)=bu^2$ in (\ref{(4.6)}) gives the integrable
equations
\begin{eqnarray}
\zeta_t&=&a(\zeta^{-3}\zeta_u)_{uu}+b,\label{(4.6a)}\\
\zeta_t&=&a(\zeta^{-3}\zeta_u)_{uu}+2bu.
\end{eqnarray}

\Example 5.
Consider the third-order equation
\begin{equation}
u_t=au_{xxx}+bu_x^n,
\label{(4.7)}
\end{equation}
which is linear for $n=0$ and $n=1$ and also integrable for
$n=2$ and $n=3$ [5,~7].
The B\"acklund transformation (\ref{(3.4)}) with
$F=au_{xxx}+bu_x^n=a\eta(\eta\eta_u)_u+b\eta^n$ and $s(t)\equiv 0$ brings
Eq.~(\ref{(4.7)}) to a nonlinear equation, which is further reduced with the
change of variable $\eta=\theta^{1/2}$ to the form
\begin{equation}
\theta_t=a\theta^{3/2}\theta_{uuu}+b(n-1)\theta^{n/2}\theta_u.
\label{(4.8)}
\end{equation}

In particular, setting $n=2$ and $n=3$ in (\ref{(4.8)}) gives the integrable
equations
\begin{eqnarray*}
\theta_t&=&a\theta^{3/2}\theta_{uuu}+b\theta\theta_u,\\
\theta_t&=&a\theta^{3/2}\theta_{uuu}+2b\theta^{3/2}\theta_u.
\end{eqnarray*}

\Example 6.
The nonlinear third-order equation
\begin{equation}
u_t=f(u)u_{xxx}
\label{(4.9)}
\end{equation}
is integrable for $f(u)=au^{3/2}$ [5, p.~535] (cf.\ Eq.~(\ref{(4.8)}) with
$b=0$) and $f(u)=au^3$ (Harry Dym equation) [5, p.~528]. The B\"acklund
transformation (\ref{(3.4)}) with $F=f(u)\eta_{xx}=f(u)\eta(\eta\eta_u)_u$ and
$s(t)\equiv 0$ reduces Eq.~(\ref{(4.9)}) to the equation
$\eta_t=\eta^2[f(u)(\eta\eta_u)_u]_{u}$, which can also be rewritten as
\begin{eqnarray*}
\theta_t=\theta^{3/2}[f(u)\theta_{uu}]_{u},\quad \  \theta=\eta^2.
\end{eqnarray*}

\Example 7. The linear fourth-order equation $u_t=au_{xxxx}$
can be reduced, using the B\"acklund
transformation (\ref{(3.4)}) with $F=a\eta_{xxx}=a\eta[\eta(\eta\eta_u)_u]_u$
and $s\equiv 0$ followed by the change of variable $\eta=\theta^{1/2}$,
to the nonlinear fourth-order equation
\begin{equation}
\theta_t=a\theta^{3/2}(\theta^{1/2}\theta_{uu})_{uu}.
\label{(4.10)}
\end{equation}

\Remark. More new integrable equations may be obtained by applying a B\"acklund
transformation of the form (\ref{(3.4)}), based on the von Mises
transformation, to each of the equations (\ref{(4.2)}) (with $f(u)=bu$),
(\ref{(4.4)}), (\ref{(4.6a)}), (\ref{(4.8)}), and (\ref{(4.10)}).

\section{Using the von Mises transformation for integro-\\differential equations}

Consider integro-differential equations of the form
\begin{equation}
u_t=F(t,u,u_x,u_{xx},\dots,u^{(n)}_x)+u_x\int^x_{x_0}
G(t,u,u_z,u_{zz},\dots,u^{(m)}_z)\,dz.
\label{(5.1)}
\end{equation}
The following brief notation is used for the arguments of~$G$:
$u^{(k)}_z=u^{(k)}_z(t,z)$, $k=0,\,1,\,\dots,\,m$.
In the special case $G=s(t)$ and $x_0=0$, Eq.~(\ref{(5.1)})
coincides with Eq.~(\ref{(3.1)}).

Dividing Eq.~(\ref{(5.1)}) by $u_x$, differentiating with respect to~$x$,
and passing to the von Mises variables (\ref{(2.2)}), we obtain
\begin{equation}
\eta_t=\eta^2\pd{}u\left(\frac F\eta\right)+\eta G,
\label{(5.2)}
\end{equation}
where $F=F(t,u,\eta,\eta\eta_u,\eta(\eta\eta_u)_u,\dots)$ and
$G=G(t,u,\eta,\eta\eta_u,\eta(\eta\eta_u)_u,\dots)$.

Equations (\ref{(5.1)}) and (\ref{(5.2)}) are related by the nonlocal
B\"acklund transformation
\begin{equation}
\begin{array}{rcl}
u_t&=&\displaystyle F(t,u,\eta,\eta_x,\dots,\eta^{(n-1)}_x)+\eta \int^x_{x_0}
  G(t,u,u_z,u_{zz},\dots,u^{(m)}_z)\,dz,\\
u_x&=&\eta.
\end{array}
\label{(5.3)}
\end{equation}

\Example. Let us choose the following functions in Eq.~(\ref{(5.1)}):
\begin{equation}
\begin{array}{rl}
F&\displaystyle=f_1(t,u)u_x+f_2(t,u)+f_3(t,u)\frac{u_{xx}}{u_x^2},\\[12pt]
G&\displaystyle=g_1(t,u)u_x+g_2(t,u)+g_3(t,u)\frac{u_{xx}}{u_x^2}.
\end{array}
\label{(5.4)}
\end{equation}
Transformation (\ref{(5.3)})--(\ref{(5.4)}) leads to the partial differential equation
$$
\eta_t=\eta^2\pd{}u\left(f_1+\frac {f_2}\eta+f_3\frac{\eta_u}{\eta^2}\right)+
g_1\eta^2+g_2\eta+g_3\eta_u.
$$
With the change of variable $\eta=1/\zeta$, it is reduced to the linear
equation
$$
\zeta_t=\pd{}u\bigl(f_3\zeta_u-f_2\zeta-f_1\bigr)+g_3\zeta_u-g_2\zeta-g_1.
$$

\Remark. The integro-differential equation (\ref{(5.1)}) can be reduced, by
dividing by~$u_x$ and differentiating with respect to~$x$, to a partial
differential equation containing a mixed derivative.

\section{A new class of nonlinear equations with mixed\\ derivatives that admit order reduction}
\label{sec:6}

By setting $G=a(t)u$ in Eq.~(\ref{(5.1)}) and substituting
\begin{equation}
w(t,x)=\int^x_{x_0}u(t,z)\,dz,
\label{(6.1)}
\end{equation}
one arrives at the following $(n+1)$st-order differential equation with a mixed derivative:
\begin{equation}
w_{tx}=a(t)ww_{xx}+F(t,w_x,w_{xx},\dots,w^{(n+1)}_x).
\label{(6.2)}
\end{equation}

Nonlinear $(n+1)$st-order equations of the form (\ref{(6.2)})
can be reduced, using substitution (\ref{(6.1)}) and applying
the B\"acklund transformation (\ref{(5.3)}) with $G=a(t)u(t,z)$,
to the $n$th-order equation
$$
\eta_t=\eta^2\pd{}u\left(\frac F\eta\right)+a(t)u\eta,\quad
F=F(t,u,\eta,\eta\eta_u,\eta(\eta\eta_u)_u,\dots).
$$

\medskip
\Example 1 (generalized Calogero equation).
The nonlinear second-order equation with a mixed derivative
\begin{equation}
w_{tx}=[f(t,w_x)+a(t)w]w_{xx}+g(t,w_x)
\label{(6.3)}
\end{equation}
is a special case of Eq.~(\ref{(6.2)}) with $F=f(t,w_x)w_{xx}+g(t,w_x)$.
Substitution (\ref{(6.1)}) brings Eq.~(\ref{(6.3)}) to an equation of the
form~(\ref{(5.1)}):
\begin{equation}
u_t=f(t,u)u_x+g(t,u)+a(t)u_x\int^x_{x_0}u(t,z)\,dz.
\label{(6.4)}
\end{equation}
The nonlocal B\"acklund transformation (\ref{(5.3)}) with $G=a(t)u(t,z)$
reduces Eq.~(\ref{(6.4)}) to the equation
$$
\eta_t=\eta^2 \pd{}u\left[f(t,u)+\frac {g(t,u)}\eta\right]+a(t)u\eta,
$$
which becomes linear after substituting $\eta=1/\zeta$.

In the special case of $a(t)=1$, $f(t,w_x)=0$, and $g(t,w_x)=g(w_x)$,
Eq.~(\ref{(6.3)}) becomes the Calogero equation [5,~20,~21].

\medbreak
\Example 2.
Consider the nonlinear third-order equation
\begin{equation}
\pdd wtx+w\pdd wxx-m\left(\pd wx\right)^{\!\!2}=
\nu\frac{\partial^3w}{\partial x^3}+q(t)\pd wx+p(t),
\label{(6.5)}
\end{equation}
which arises in hydrodynamics and derives from the Navier--Stokes equations
[5,~22--24]; $w$ is one of the fluid velocity components
and $\nu$ is the kinematic viscosity. With $a(t)=-1$, $F=\nu
u_{xx}+mu^2+q(t)u+q(t)$, and $u=w_x$, the above transformations reduce Eq.~(\ref{(6.5)})
to a second-order equation representable in the form
\begin{equation}
\pd\eta t+[mu ^2+q(t)u +p(t)]\pd\eta u =[(2m-1)u +q(t)]\eta+\nu\eta^2\pdd\eta uu.
\label{(6.6)}
\end{equation}

Note that in the degenerate case of inviscid fluid ($\nu=0$),
the original nonlinear equation~(\ref{(6.5)}) reduces to a first-order linear
equation, Eq.~(\ref{(6.6)}) with $\nu=0$.

\section{Generalization of the von Mises transformation to the case of three independent variables}

The von Mises transformation can be generalized to the cases of three or more
independent variables. For example, let us consider the $n$th-order equation
with three independent variables
\begin{equation}
F(t,x,u,u_y,u_{yy},\ldots,u^{(n)}_y,u_yu_{ty}-u_tu_{yy},u_yu_{xy}-u_xu_{yy})=0.
\label{(7.1)}
\end{equation}
The von Mises transformation
\begin{equation}
t, \ x, \ y, \ u=u(t,x,y)\quad \Longrightarrow\quad
t, \ x, \ u, \ \eta=\eta(t,x,u),\ \ \hbox{where}\ \ \eta=\pd uy,
\label{(7.2)}
\end{equation}
reduces the order of Eq.~(\ref{(7.1)}) and brings it to the form
\begin{equation}
F(t,x,u,\eta,\eta\eta_u,\ldots,\eta^{(n-1)}_y,\eta\eta_t,\eta\eta_x)=0.
\label{(7.3)}
\end{equation}
where the $y$-derivatives are calculated as
$\eta^{(k)}_y=\eta\bigl(\eta^{(k-1)}_y\bigr)_u$.

\medskip
\Example.
Consider the Prandtl system
\begin{equation}
\begin{array}{rl}
u_t+uu_x+vu_y&=\nu u_{yy}+f(t,x),\\
u_x+v_y&=0.
\end{array}
\label{(7.4)}
\end{equation}
It describes a plane unsteady boundary layer with pressure
gradient; $u$ and $v$ are the fluid velocity components [3--5].
Let us divide the first equation by~$u_y$, then differentiate with respect to~$y$,
and eliminate~$v_y$ using the second equation in~(\ref{(7.4)})
to obtain a nonlinear third-order equation of the form~(\ref{(7.1)}):
\begin{equation}
u_yu_{ty}-u_tu_{yy}+u(u_yu_{xy}-u_xu_{yy})=\nu(u_yu_{yyy}-u_{yy}^2)-f(t,x)u_{yy}.
\label{(7.5)}
\end{equation}
Applying the von Mises transformation (\ref{(7.2)}) to Eq.~(\ref{(7.5)})
yields the second-order equation
\begin{equation}
\pd\eta t+u \pd\eta x+f(t,x)\pd\eta u=\nu\eta^2\pdd\eta uu.
\label{(7.6)}
\end{equation}
Once a solution to this equation has been found, the second velocity component~$v$
is found immediately from the first equation of~(\ref{(7.4)}) without integrating.

Note that Eq.~(\ref{(7.6)}) can be reduced, with the change of variable $\eta=1/\zeta$,
to the nonlinear heat equation
\begin{equation}
\pd\zeta t+u \pd\zeta x+f(t,x)\pd\zeta u=\nu\pd{}u\left(\frac 1{\zeta^2}\pd\zeta u\right).
\label{(7.6a)}
\end{equation}

\eqnitem 1.
Consider the special case $f(t,x)=f(t)$.
We look for exact solutions to Eq.~(\ref{(7.6a)}) in the form
$$
\zeta=Z(\xi,\tau),\quad \xi=x-u t+\tint tf(t)\,dt,\quad \tau=\frac 13t^3.
$$
We arrive at the integrable equation
\begin{equation}
\pd Z\tau=\nu\pd{}\xi\left(\frac 1{Z^2}\pd Z\xi\right),
\label{(7.7)}
\end{equation}
which can be reduced to the linear heat equation [19] (see also~[5]).

\eqnitem 2.
Consider the more general case $f(t,x)=f(t)x+g(t)$.
We look for exact solutions of the special form
$$
\zeta=Z(\xi,\tau),\quad
\xi=\varphi(t)x+\psi(t)u+\theta(t),\quad
\tau=\tint \psi^2(t)\,dt,
$$
where $\varphi=\varphi(t)$, $\psi=\psi(t)$, and $\theta=\theta(t)$
are determined by the linear system of ordinary differential equations
$$
\varphi'_t+f\psi=0,\quad
\psi'_t+\varphi=0,\quad
\theta'_t+g\psi=0.
$$
As a result, we arrive at the integrable equation (\ref{(7.7)}).

\section{Conclusion}

We have presented wide classes of nonlinear mathematical physics equations that
admit order reduction through the use of the von Mises transformation, with
the unknown function taken as the new independent variable and an
appropriate partial derivative taken as the new dependent variable.
RF-pairs and associated B\"acklund transformations are constructed for
evolution equations of general form (special cases of which are
Burgers and Korteweg--de Vries type equations as well as many
other nonlinear equations of mathematical physics). The results obtained
are used for order reduction and constructing exact solutions of
hydrodynamics equations. A generalized Calogero equation and a number of
other new integrable nonlinear equations are considered.

The von Mises and B\"acklind transformations described in the paper
can be used to construct other new integrable nonlinear differential and
integro-differential equations
of mathematical physics.

\section*{Acknowledgments}
The work was carried out under partial financial support of the Russian
Foundation for Basic Research (grants No.~\hbox{08-01-00553},
No.~\hbox{08-08-00530} and No.~\hbox{09-01-00343}).

\end{document}